\documentclass[aps,noshowpacs,superscriptaddress,nofootinbib,preprint,11pt]{revtex4}
\usepackage{graphicx}
\usepackage{array}
\usepackage{placeins}

\usepackage{amsmath, amssymb, graphics, setspace, hyperref}
\newcommand{\mathsym}[1]{{}}
\newcommand{\unicode}[1]{{}}
\allowdisplaybreaks

\newcommand{\be}{\begin{equation}}
\newcommand{\ee}{\end{equation}}

\newcommand{\calO}{{\cal{O}}}

\newcommand {\E}{\nabla\phi}
\newcommand {\bea}{\begin{eqnarray}}
\newcommand {\eea}{\end{eqnarray}}

\def\beq{\begin{eqnarray}}\def\eeq{\end{eqnarray}}
\def\be{\begin{equation}}\def\ee{\end{equation}}
\def\o{\omega}

\def\o{\omega}

\def\x{{\bf x}}

\renewcommand{\vec}[1]{\mathbf{#1}}

\numberwithin{equation}{section}

\begin{document}

\title{Viscosity for Anisotropic Reissner Nordstr{\"o}m Blackbranes}
\author{Soumangsu Chakraborty}
\email{soumangsubhusan.chakraborty@mail.huji.ac.il}
\affiliation{Racah Institute of Physics, Hebrew University of Jerusalem, Jerusalem, 91904 Israel}
\author{Rickmoy Samanta}
\email{rickmoysamanta@gmail.com}
\affiliation{Department of Physics, Bar Ilan University, Ramat Gan, 52900 Israel}

\begin{abstract}
We investigate the behavior of shear viscosity in the presence of small anisotropy and a finite chemical potential. First, we construct an anisotropic  Reissner Nordstr{\"o}m blackbrane in 5 dimensions in a simple Einstein-Maxwell theory with a small linear dilaton. This solution is characterized by three mass scales : anisotropy $\rho$, temperature T and chemical potential $\mu$. We find this solution upto second order in the dilaton anisotropy parameter $\rho$. This blackbrane solution corresponds to an anisotropic phase where the anisotropy is small compared to the temperature and chemical potential.  We find that in this anisotropic phase, some components of the anisotropic shear viscosity tensor, which are spin one with respect to the residual symmetry after breaking rotational invariance,  violates the KSS bound (${\eta \over s}\ge {1 \over 4 \pi} $) proposed by Kovtun, Son and Starinets. We identify the regions of the parameter space where these violations are significant. We carry out a similar analysis in 4 dimensions and find similar violation of the KSS bound for the spin one components to demonstrate the generality of the result. Our results are particularly relevant in the context of strongly coupled systems found in nature. We also provide an intuitive understanding of the results using dimensional reduction and a Boltzmann calculation in a weakly coupled version of a similar system.  The Boltzmann analysis performed in a system of weakly interacting particles in a linear potential also shows that components of the viscosity tensor may be reduced. It is intriguing that the Boltzmann analysis also predicts the corrections to be negative and that too in a manner similar to the anisotropic strongly coupled  theories with smooth gravity duals.

\end{abstract}
\vskip .5 true cm

\maketitle

\section{Introduction}

The gauge-gravity correspondence has been very useful in the computation of
transport coefficients in strongly coupled field theories that admit smooth gravitational duals.
Such computations revealed  \cite{Policastro:2001yc,Kovtun:2003wp,Kovtun:2004de} that for systems having a dual gravity description, the ratio of  shear viscosity, $\eta$, to the entropy density, $s$, becomes
 \be
 \label{kss}
 {\eta \over s}={1\over 4 \pi}.
 \ee
It was proposed by KSS ( Kovtun, Son and Starinets) that this value may be a bound and the ratio cannot be smaller than ${1 \over {4 \pi}}$. It is now known that this is not true \cite{Kats:2007mq,Buchel:2008vz,Sinha:2009ev,Cremonini:2011iq}, see also \cite{Basu:2011tt,Bhattacharyya:2014wfa}, but in most of the controlled counter-examples the bound is violated at best  by a numerical factor, and not in a parametric manner. Significant violations of the bound lead to unphysical situations, e.g., to causality violations, for example, see \cite{Brigante:2007nu,Brigante:2008gz}. 
However, see \cite{Cohen:2007qr} for discussion  of  a violation  of the bound in metastable states and also \cite{Buchel:2010wf} for a discussion of violations in a superfluid phase described by higher derivative gravity.\\
The above results hold for isotropic situations.
Over the last few years, many gravitational backgrounds featuring anisotropy (see
\cite{Landsteiner:2007bd,Azeyanagi:2009pr,Natsuume:2010ky,Erdmenger:2010xm,
Basu:2011tt,Erdmenger:2011tj,Mateos:2011ix,Mateos:2011tv,Iizuka:2012wt,Cheng:2014qia,Roychowdhury:2015fxf})has been studied and using the AdS/CFT correspondence , the shear viscosity has also been computed in some of these anisotropic phases.
(see \cite{Rebhan:2011vd, Polchinski:2012nh} and
\cite{Giataganas:2012zy,Mamo:2012sy,Jain:2014vka,Critelli:2014kra,Ge:2014aza, Bhattacharyya:2014wfa}). In such anisotropic examples it has been found that different components of the shear viscosity tensor can behave differently and some components in fact parametrically violate the KSS bound.\\
For example, one can consider a simple setup consisting of Einstein gravity and a massless dilaton with a linear profile in one of the coordinates, (say $\phi=\rho z $) that breaks rotational invariance but the stress tensor remains translation invariant. Such a system was considered in ~\cite{Jain:2014vka} with the action 
\begin{equation}
S = \frac{1}{16\pi G} \int d^5 x \sqrt{g}~ [R+12\Lambda -
\frac{1}{2}\partial_\mu \phi\partial^\mu\phi]\;~\label{eq:5dlag}
\end{equation}
and the dilaton varies linearly along one of the spatial directions ie. $\phi= \rho z$. It was shown in ~\cite{Jain:2014vka} that in this system, at a finite temperature 
$T$, the components of the shear viscosity tensor which
 are spin 1 with respect to the surviving Lorentz symmetry in the boundary theory \footnote{ The residual Lorentz symmetry exists at zero temperature, after breaking the rotational symmetry.
Fluid mechanics is then governed by the dynamics of the goldstone modes corresponding to the boost symmetries of this Lorentz group broken at finite temperature.}( eg. $\eta_{xzxz}$ which we abbreviate as $\eta_{xz}$) parametrically violate the KSS bound. For example, in the low anisotropy regime ($\rho/T \ll 1$):
\begin{equation}
\frac{\eta_{xz}}{s}=
\frac{1}{4\pi}-\frac{\rho^2 \log 2}{16 \pi^3 T^2}
+
\frac{(6-\pi^2+54 (\log 2)^2)\rho^4}{2304\pi^5 T^4}
+
{\cal{O}}\bigg[\bigg(\frac{\rho}{T}\bigg)^6\bigg]~\label{eq:eta_low}\;.
\end{equation}

In the limit of extreme anisotropy ($\rho/T \gg 1$),
\begin{equation}
{\eta_{xz}}/{s}\rightarrow ({1}/{4\pi}) ({32\pi^2 T^2}/{3\rho^2})\,
\end{equation}
and hence becomes parametrically small~\cite{Jain:2014vka} as ${T\over \rho}\rightarrow 0$.\\
However the $\eta_{xy}$ component which is spin 2 with respect to the residual symmetry remained unchanged from the answer in the isotropic case. 
\begin{eqnarray}
\frac{\eta_{xy}} {s}= \frac{1}{4\pi}.
\end{eqnarray}\\
In  \cite{Jain:2015txa} many examples where anisotropic phases arise were
studied and in all of these cases it was found that the spin $1$ components of
the viscosity can become parametrically small, in units of the entropy
density, when the anisotropy becomes sufficiently large compared to the
temperature.  It was shown in  \cite{Jain:2015txa}  that for all situations where the force responsible for breaking rotational symmetry is translation invariant and a residual Lorentz symmetry survives\footnote{Our situation is different from \cite{Hartnoll:2016tri} which considers isotropic backgrounds by turning on massless scalars along all the spatial directions. See also \cite{Alberte:2016xja,Ling:2016ien,Ling:2016yxy} for interesting new directions from \cite{Hartnoll:2016tri}.} in the boundary theory after breaking of rotational invariance, there exists a general formula for the spin 1 shear viscosity components (in units of entropy density) in terms of metric components evaluated at the horizon.
The general formula can be presented as follows: let $z$ be the field theory direction along which   a spatially constant driving force is turned on breaking rotational symmetry and $x$ be a direction   along which the boost symmetry is left unbroken, 
then the viscosity component $\eta_{xz}$ is given by
\begin{equation}
\label{mresaa}{\eta_{xz}\over s} = {1\over 4\pi} 
{{ g}_{xx}\over { g}_{zz}}~\Big{|}_{u=u_{h}}, 
\end{equation}
where ${ g}_{xx}|_{u=u_{h}}, {g}_{zz}|_{u=u_{h}}$ refer to the components of
the background metric evaluated at the horizon. This  result is true for all the anisotropic situations studied in \cite{Jain:2015txa}. This result was first derived in an anisotropic axion-dilaton system considered in \cite{Rebhan:2011vd}.\\
In the isotropic situation, the metric components $g_{xx}$ and $g_{zz}$ are the same and we recover the result $\frac{1}{4 \pi}$ of KSS. However in anisotropic situations, these metric components can behave very differently and thus leads to the parametric violations of the KSS bound.\\
All such anisotropic situations involve two scales : one is the anisotropy parameter and the other is the temperature. One may ask whether the violations of the KSS bound in such anisotropic situations exist in the presence of a third scale - a finite chemical potential($\mu$). Using the general formula Eq.\eqref{mresaa}  one can compute the spin 1 components of shear viscosity in the presence of a  chemical potential. One can thus easily check the KSS bound for anisotropic systems in the presence of a finite chemical potential. This is the question we try to address in this paper. The question becomes interesting  because the system we consider has many similarities with the strongly coupled systems found in nature, eg ultracold atoms at unitarity, which features two scales T and $\mu$, along which we introduce a third scale ie. anisotropy. This may be implemented easily in ultracold atom setups by considering anisotropic traps. ( see \cite{Samanta:2016lsh} for details ). \\
We first consider the well known charged Reissner Nordstr{\"o}m(RN) blackbrane in a simple Einstein Maxwell theory in 5 dimensions. We turn on a small dilaton breaking the rotational symmetry along one of the field theory directions. We then perturbatively compute the metric corrections upto  second order in the dilaton anisotropy parameter, demanding that the geometry asymptotes to $AdS$. This blackbrane solution corresponds to an anisotropic phase where the anisotropy is small compared to the temperature and chemical potential. We then use the general formula Eq.\eqref{mresaa} to compute the shear viscosity to entropy density ratio for the spin 1 components in this low anisotropy regime in the presence of a finite chemical potential.\\
We find that in 5 dimensions in the regime $ \rho \ll (\mu, T)$,
 \begin{eqnarray}
  {\eta \over s}=\frac{1}{4 \pi }- {\rho ^2  \over \mu^2}\frac{\sqrt{3}f^2 \tanh ^{-1}\left(\frac{\sqrt{12 f^2 +9}}{9}\right)}{8 \pi  \sqrt{4  f^2+3}} +O(\rho^4),
   \label{5dvisintro}
  \end{eqnarray}
where $f=-3 \pi {T \over \mu} +\sqrt{3}\sqrt{ 2 + 3 \pi^2 {T^2 \over \mu^2}  }$.\\
 We find that the corrections to ${\eta \over s}$ due to anisotropy is negative even in the presence of a finite chemical potential. We find similar results in 4D as well. This is the main result of this paper. Similar results were also found in \cite{Ge:2014aza} where it was shown that these spin 1 components of shear viscosity violated the bound in the Einstein-Maxwell-dilaton-axion theory. Surprisingly, a Boltzmann calculation in a system of weakly interacting particles in a linear potential also predicts the corrections to go in a manner similar to Eq.\eqref{5dvisintro}. (see Appendix B of \cite{Samanta:2016pic} for a detailed calculation). Motivated by these interesting results, an experiment  has also been proposed in \cite{Samanta:2016lsh} to measure such spin one components of shear viscosity in the unitary fermi gases.\\
One can also try to gain an intuitive understanding of these results along the lines of \cite{Jain:2015txa} in order to understand the  parametric violation of the ${\eta \over s}$ bound in such anisotropic situations. The key idea is that  dimensional reduction maps the spin 1 gravitational shear modes under consideration to massive gauge fields in the lower dimensional theory and thus, the shear viscosity gets related to conductivity in the dimensionally reduced theory. It is well known in several AdS/CFT examples that conductivity can become very small. This helps us understand the reduction of spin 1 viscosity components which gets related to conductivity in the dimensionally reduced description. In Appendix \ref{condres}, using the same idea, we show that the formula for electrical conductivity to susceptibility ratio (${\sigma \over \chi}$) by Kovtun and Ritz in  \cite{Kovtun:2008kx} may be looked upon as a consequence of the shear viscosity over entropy density formula in a higher dimensional theory.\\

The paper is organized as follows : We present the action in 5D and conventions in Sec.\ref{setup5d} and the charged blackbrane solution in Sec.\ref{rn5d}. We next turn on a small linear dilaton and perturbatively compute the metric corrections upto second order in the dilaton anisotropy parameter in Sec.\ref{pert5d}. The spin 1 viscosity to entropy ratio is computed using the general formula Eq.\eqref{mresaa} in Sec.\ref{vis5d}. We carry out a similar procedure in 4 dimensions in Sec.\ref{setup4d}. We compare these results with that of a system of weakly interacting charged particles in an external field in Sec.\ref{comp}. Appendix \ref{condres} contains details of how these spin one viscosity components are related to conductivity in a lower dimensional theory. \\

\section{Anisotropic solution in 5D dilaton gravity system in the presence of a finite chemical potential}
\label{setup5d}

We consider a 5 dimensional gravity system with a massless scalar field $\phi$, a timelike massless one form $U(1)$ gauge field $\mathcal{A}$ and a constant cosmological constant $\Lambda$ with action,
\begin{eqnarray}
S_{bulk}=\frac{1}{16\pi G}\int d^5 x\sqrt{-g}\left(R+12\Lambda-\frac{1}{2}\left(\partial\phi\right)^2-\frac{1}{4}F_{\mu\nu}F^{\mu\nu}\right),
\end{eqnarray}
where the two form field strength is given by $F=d\mathcal{A}$ and $G$ is the Newton's constant in 5 dimension. \\
The equations of motion of this system are given by:
\begin{eqnarray}
R_{\mu\nu} -\frac{1}{2} g_{\mu \nu}R&=&\frac{1}{2}\partial_{\mu}\phi\partial_{\nu}\phi+\frac{1}{2} F_{\mu\alpha}F_{\nu}^{\alpha}+\frac{1}{2}  g_{\mu \nu}\left(12 \Lambda -\frac{1}{4}F^2 -\frac{1}{2}(\partial\phi)^2\right),\label{einsteins eqn}\\
\Box\phi&=&0,\label{dilaton equation}\\
\partial_\nu(\sqrt{-g}F^{\mu\nu})&=&0.\label{gauge field equation}
\end{eqnarray}
 In the absence of the gauge field and dilaton, the above action has the $AdS_5$  solution given by
\begin{eqnarray}\label{AdS5 solution}
ds^2=L^2\left[-r^2dt^2+\frac{dr^2}{r^2}+r^2\left(dx^2+dy^2+dz^2\right)\right]
\end{eqnarray} 
where $L$ the radius of the $AdS$ space which is related to $\Lambda$ as 
\begin{eqnarray}
\Lambda=\frac{1}{L^2}.
\end{eqnarray}
In generalized Gaussian null coordinates the above $AdS_5$ solution takes the form
\begin{eqnarray}\label{AdS5 in gaussian null coordinates}
ds^2=\frac{L^2}{u^2}\left[-dv^2-2dudv+dx^2+dy^2+dz^2\right].
\end{eqnarray}
The following change of coordinates 
\begin{eqnarray}
v=t-u,\ \ \ \ \ u=\frac{1}{r}
\end{eqnarray}
in the above $AdS_5$ solution in Eq.\eqref{AdS5 in gaussian null coordinates} gives back the solution in Eq.\eqref{AdS5 solution}.\\
Another interesting solution is obtained by breaking rotational  invariance in the three spacelike directions $x,y \text{ and } z$. One way to achieve this is by turning on a dilaton (with vanishing gauge field) of the form:
\begin{eqnarray}
\phi=\rho z
\end{eqnarray}
where $\rho$ is an additional scale that breaks rotational invariance. In \cite{Jain:2014vka} the authors have argued that the backreaction of the dilaton will preserve rotational invariance in the $x, y$ direction. Moreover since the dilaton stress tensor is independent of $t,x,y,z$, the metric will be invariant under translation in these directions. This gives rise to a class of $AdS$ black brane solutions which are homogeneous but anisotropic. The details of this solution is discussed in \cite{Jain:2014vka}.\\ 
Our aim is to make this problem even more interesting (and somewhat more closer to situations arising in nature) by turning on a timelike gauge field $\mathcal{A}=\mathcal{A}_v(u)dv$ in the bulk which corresponds to the chemical potential in the dual field theory residing on the conformal boundary of the asymptotically $AdS$ spacetime. Before we start to analyze out system in details let us first construct a metric ansatz. In generalized Gaussian null coordinates one can propose a static regular homogeneous metric ansatz that preserves rotational invariance only in the $x,y$ directions of the form
\begin{eqnarray}\label{metric ansatz}
ds^2&=&\frac{L^2}{u^2}\left[-g(u)dv^2-2dvdu+e^{A(u)-B(u)}\left(dx^2+dy^2\right)+e^{A(u)+B(u)}dz^2\right],
\end{eqnarray}
where $B(u)$ generates the anisotropy of the metric ansatz with $g(u)>0$ outside the horizon. Setting $g(u)=1$ and $A(u)=B(u)=0$ give back the $AdS_5$ solution in Eq.\eqref{AdS5 in gaussian null coordinates} as expected. \\
There are quite a few added advantages in working with Gaussian null coordinates over the Schwarzschild like coordinates used in Eq.\eqref{AdS5 solution} (see \cite{Iizuka:2012wt} for details). The horizon to this blackbrane is represented as $u=$constant null hypersurface where $g(u)$ goes to zero. The determinant of the metric being non-zero at the horizon implies that the metric is smooth at the horizons with no coordinate singularity. The equations of motion being second order partial differential equations, it is quite natural to assume that the functions $A(u),B(u),g(u)$ are smooth and atleast twice differentiable.
On substitution of the metric ansatz Eq.\eqref{metric ansatz} in the equations of motion \eqref{einsteins eqn}, \eqref{dilaton equation} and \eqref{gauge field equation} gives
\begin{eqnarray}
&&e^{A(u)} \bigg[u g'(u) \left(3 u A'(u)-u B'(u)-6\right)+2
   g(u) \Big[u \Big(u \left(3 A''(u)-B''(u)+B'(u)^2\right)\nonumber \\
   &&\ \ \ \ \ -A'(u) \left(2 u B'(u)+9\right)+3 u
   A'(u)^2+3 B'(u)\Big)+12\Big]+u^4 \mathcal{A}_v'(u)^2-24\bigg]\nonumber \\
  &&\ \ \ \ \  +\rho ^2 u^2e^{-B(u)}=0,\label{e o m 1}\\
  &&u \bigg[4 \left(u A'(u)-3\right) g'(u)+u^3 \left(-\mathcal{A}_v'(u)^2\right)+2 u g''(u)\bigg]+g(u)
   \bigg[u \bigg(u \left(4 A''(u)+B'(u)^2\right)\nonumber \\
   &&\ \ \ \ \ +3 u A'(u)^2-12 A'(u)\bigg)+24\bigg]-24=0,\label{e o m 2}\\   
  && e^{A(u)} \bigg[4 u g'(u) \big(u \left(A'(u)-B'(u)\right)-3\big)+g(u) \big(u \big(4 u\big(A''(u)-B''(u)\big)-6 A'(u) \big(u B'(u)+2\big)\nonumber \\
   &&+3 u A'(u)^2+3 u B'(u)^2+12
   B'(u)\big)+24\big)+u^4 \big(-\mathcal{A}_v'(u)^2\big)+2 u^2 g''(u)-24\bigg]\nonumber\\
   && \ \ \ \ \ -\rho ^2 u^2
   e^{-B(u)}=0,\label{e o m 3}\\
  &&2 \left(3 A''(u)-B''(u)\right)-2 A'(u) B'(u)+3 A'(u)^2+3 B'(u)^2=0,\label{e o m 4}\\
   &&   \mathcal{A}_v'(u) \left(3 u A'(u)-u B'(u)-2\right)+2 u \mathcal{A}_v''(u)=0.\label{e o m 5}
\end{eqnarray} 
One can easily verify that  $A(u)=B(u)=0$ and $g(u)=1$ with vanishing dilaton and gauge field indeed satisfy the above equations of motion. This, as stated earlier gives back the $AdS_5$ solution of pure gravity in 5 dimension with cosmological constant.

\subsection{5D Reissner Nordstr{\"o}m blackbrane solution}
\label{rn5d}

To recover the RN blackbrane solution we switch off the dilaton field and set $A(u)=B(u)=0$. This gives rise to the following equation of motion;
\begin{eqnarray}
&&u^4 \mathcal{A}_v'(u)^2-6 \left(u g'(u)+4\right)+24 g(u)=0,\\
&&u^4 \mathcal{A}_v'(u)^2-2 u^2 g''(u)+12 u g'(u)-24 g(u)+24=0,\\
&&u \mathcal{A}_v''(u)-\mathcal{A}_v'(u)=0.
\end{eqnarray}
Demanding the gauge field to vanish at the outer horizon($u_h$), the above list of differential equations can be solved as;
\begin{eqnarray}
\mathcal{A}_v(u)&=&\frac{\mu}{u_h^2}  \left(u_h^2-u^2\right),\\
g(u)&=&\frac{1}{3} \left[{\frac{\mu}{u_h^2}} ^2 u^6-\frac{u^4 \left(\mu ^2 u_h^2+3\right)}{u_h^4}+3\right].
\end{eqnarray}
One can factorize $g(u)$ as;
\begin{eqnarray}
g(u)=\frac{(u-u_h) (u+u_h) \left(\mu ^2 u^4 -3 u^2-3 u_h^2\right)}{3u_h^4}.
\end{eqnarray}
Hence the temperature associated to the outer horizon is given by
\begin{eqnarray}
T=\frac{6-\mu ^2 u_h^2}{6 \pi  u_h}
\end{eqnarray}
In these coordinates, the boundary is at $u=0$ and the inner horizon is at $\frac{u_h}{\xi}$ where $0<\xi<1$. Thus, $\frac{u_h}{\xi}$ is a root of the horizon function i.e.
\begin{equation}
\left(\mu ^2\frac{u_h^4 }{ \xi^4} -3 u_h^2-3 \frac{u_h^2}{\xi^2}\right)=0.
\end{equation} 
Solving $\xi$ from the above equation gives the expression for the inner horizon($u_{inner}$)  as;
\begin{eqnarray}\label{innhor}
u_{inner}=\frac{\sqrt{6} u_h}{\sqrt{-3+\sqrt{9+12u_h^2 \mu^2}}}
\end{eqnarray}
Demanding $0<\xi<1$ and $T>0$ we get $u_h>0$ and $-\frac{\sqrt{6}}{u_h}<\mu<\frac{\sqrt{6}}{u_h}$.
\subsection{Dilaton perturbation in Reissner Nordstr{\"o}m blackbrane}
\label{pert5d}
Now we switch on the dilaton field $\phi(z)=\rho z$ and consider perturbation around the RN blackbrane solution. Let us expand $A(u),B(u)$ and $g(u)$ in power series of $\rho$:
\begin{eqnarray}
A(u)&=&0+\rho^2a_2(u)+\mathcal{O}(\rho^4),\\
B(u)&=&0+\rho^2b_2(u)+\mathcal{O}(\rho^4),\\
g(u)&=&\frac{(u-u_h) (u+u_h) \left(\mu ^2 u^4 -3 u^2-3 u_h^2\right)}{3
   u_h^4}+\rho^2g_2(u)+\mathcal{O}(\rho^4),\\
   \mathcal{A}_v(u)&=&{\mu \over {u_h^2}}  \left(u_h^2-u^2\right)+\mathcal{O}(\rho^4).
\end{eqnarray}
Substituting these in the equations of motion \eqref{e o m 1}-\eqref{e o m 5} we can solve for the functions $a_2(u),b_2(u)$ and $g_2(u)$ as
\begin{eqnarray}
a_2(u)&=&-\frac{b_2(u)}{3}+c_g,\label{a2sol}\\
b_2(u)&=&c_2-\frac{\sqrt{3} u_h^2 \tanh ^{-1}\left(\frac{2 \mu ^2 u^2 -3}{\sqrt{12 \mu ^2
    u_h^2+9}}\right)}{4 \sqrt{4 \mu ^2  u_h^2+3}},\label{b2sol}\\
    g_2(u)&=&-\frac{u^2}{12}+u^4c_m,\label{g2sol}
\end{eqnarray}
where $c_g,c_2$ and $c_m$ are constants of integration to be determined by appropriate boundary condition.\\
Let us briefly comment on how we fixed one of the constants in the solution $b_2(u)$. The general solution is 
\begin{equation}
b_2(u)=c_2-\frac{\chi_1 (u)-\chi_2(u)}{16(-6+u_h^2 \mu^2) \sqrt{4 \mu^2  u_h^2+3}}
\end{equation}
where the functions $\chi_1(u)$ and $\chi_2(u)$ are given by 
\begin{eqnarray}
\chi_1(u)&=&-2 \sqrt{3}(-3 u_h^2 +2 u_h^4 \mu^2 +12 c_1) \tanh ^{-1}\left(\frac{2 \mu ^2 u^2  -3}{\sqrt{12 \mu^2  u_h^2+9}}\right)\text{ and }\\ 
\chi_2(u)&=&\sqrt{3 +4 u_h^2 \mu^2} ~(3 u_h^2+4c_1)\left(2 \log (u^2-u_h^2)-\log(-3 u^2 -3 u_h^2 +u^4  \mu^2)\right).
\end{eqnarray}
The derivative of this general solution $b_2(u)$ has the form
\begin{equation}
\partial_u b_2(u)=\frac{4 c_1 u^3 +3 u u_h^4}{4(u-u_h)(u+u_h)(-3u^2-3u_h^2+u^4  \mu^2)}
\end{equation}
Demanding regularity of the solution at the horizon, the numerator must vanish at $u=u_h$. This fixes the value of $c_1$ to be $- \frac{3}{4} u_h^2$ and we get Eq.\eqref{b2sol}.\\
 Substituting Eqs.~\eqref{a2sol},\eqref{b2sol} and \eqref{g2sol} in the metric ansatz \eqref{metric ansatz} gives
\begin{eqnarray}
g_{vv}&=&-\frac{1}{u^2}\left[\frac{(u-u_h) (u+u_h) \left(\mu ^2 u^4 -3 u^2-3 u_h^2\right)}{3 u_h^4}+\rho^2\left(-\frac{u^2}{12}+u^4c_m\right)\right],\\
g_{xx}&=&g_{yy}=\frac{1}{u^2}\exp\Bigg[{\rho^2 \over 6}\left(-4 c_2 +6 c_g -\frac{\sqrt{3} u_h^2 \tanh ^{-1}\left(\frac{2 \mu ^2 u^2 -3}{\sqrt{12 \mu ^2
   u_h^2+9}}\right)}{ \sqrt{4 \mu ^2 u_h^2+3}}\right)\Bigg]\nonumber \\
g_{zz}&=&\frac{1}{u^2}\exp\Bigg[{\rho^2 \over 3}\left(4 c_2 + 3 c_g +\frac{\sqrt{3} u_h^2 \tanh ^{-1}\left(\frac{2 \mu ^2 u^2 -3}{\sqrt{12 \mu ^2
   u_h^2+9}}\right)}{ \sqrt{4 \mu ^2 u_h^2+3}}\right) \Bigg]\nonumber \\
\end{eqnarray}
Asymptotically (i.e. $u\to 0$) we expect to get $AdS_5$. Thus expanding $g_{xx}$ and $g_{zz}$ about the asymptotic boundary $u=0$ and setting $g_{xx}=g_{zz}\sim \frac{1}{u^2}$ one can calculate the constants $c_2$ and $c_g$ as
\begin{eqnarray}
c_2&=&\frac{\sqrt{3} u_h^2 \tanh ^{-1}\left(\frac{3}{\sqrt{12 \mu ^2 u_h^2+9}}\right)}{4 \sqrt{4 \mu ^2 u_h^2+3}},\\
c_g&=&0.
\end{eqnarray} 
\subsection{Spin 1 viscosity component in 5D}
\label{vis5d}
It was shown in \cite{Jain:2015txa} that in all situations where rotational invariance is broken by a spatially constant driving force, there exists a general formula for the spin 1 shear viscosity components in terms of the ratio of appropriate metric components evaluated at the horizon. These components are spin 1 w.r.t the residual symmetry after breaking of rotational invariance. In the dual field theory, the spin 1 viscosity (in units of the entropy density $s$) component is given by;
\begin{eqnarray}
{\eta \over s} =\frac{1}{4\pi}\frac{g_{xx}}{g_{zz}} 
\end{eqnarray}  
 evaluated at the horizon. Here $z$ is the direction along which we have the translation invariant driving force and $x$ is the direction along which boost symmetry is left unbroken. Now we evaluate $\eta/s$ at $u_h$ 
 and then expand the result in power series of $\rho$ around $\rho=0$. This gives;
  \begin{eqnarray}
  \frac{\eta}{s}=\frac{1}{4 \pi }+\rho ^2 \left(-\frac{c_2}{2 \pi }-\frac{\sqrt{3} u_h^2 \tanh ^{-1}\left(\frac{2 \mu
   ^2 u_h^2-3}{\sqrt{3} \sqrt{4 \mu ^2 u_h^2+3}}\right)}{8 \pi  \sqrt{4 \mu ^2
   u_h^2+3}}\right)+O(\rho^4)
  \end{eqnarray}
  Substituting  $c_2$ in the above expression of $\eta/s$ gives\footnote{Note that our perturbation theory breaks down in the extremal limit. See \cite{Iizuka:2012wt} where a similar issue exists in 4 dimensions and \cite{Cheng:2014qia} in the dilaton-axion system.}
  \begin{eqnarray}
  {\eta \over s}=\frac{1}{4 \pi }-\frac{\sqrt{3} \rho ^2 u_h^2 \left(\tanh ^{-1}\left(\frac{3}{\sqrt{12 \mu ^2
   u_h^2+9}}\right)+\tanh ^{-1}\left(\frac{2 \mu ^2 u_h^2-3}{\sqrt{12 \mu ^2
   u_h^2+9}}\right)\right)}{8 \pi  \sqrt{4 \mu ^2 u_h^2+3}}+O(\rho^4),
   \label{5dv}
  \end{eqnarray}
where $T$ and $\mu$ are related to $u_h$ as
\begin{eqnarray}
T=\frac{6-\mu ^2 u_h^2}{6 \pi  u_h}.
\label{t5d}
\end{eqnarray}  
One can also solve for $u_h$ in terms of $\mu$ and T using the relation Eq.\eqref{t5d} and plugging it back in Eq.\eqref{5dv} to get ${\eta \over s}$ in terms of the parameters T and $\mu$. We get $u_h=\frac{-3 \pi {T \over \mu} +\sqrt{3}\sqrt{ 2 + 3 \pi^2 {T^2 \over \mu^2}  }}{\mu}$ and thus in the regime  $ \rho \ll (\mu, T)$
  \begin{eqnarray}
  {\eta \over s}=\frac{1}{4 \pi }- {\rho ^2  \over \mu^2}\frac{\sqrt{3} f^2 \left(\tanh ^{-1}\left(\frac{3}{\sqrt{12 
  f^2+9}}\right)+\tanh ^{-1}\left(\frac{2  f^2-3}{\sqrt{12 
   f^2+9}}\right)\right)}{8 \pi  \sqrt{4  f^2+3}} +O(\rho^4),
   \label{5dvis}
  \end{eqnarray}
where $f=-3 \pi {T \over \mu} +\sqrt{3}\sqrt{ 2 + 3 \pi^2 {T^2 \over \mu^2}  }$. \\
The inverse tanhyperbolic functions may be combined to give a more compact expression which we presented in the introduction, Eq.\ref{5dvisintro},
\begin{eqnarray}
  {\eta \over s}=\frac{1}{4 \pi }- {\rho ^2  \over \mu^2}\frac{\sqrt{3}f^2 \tanh ^{-1}\left(\frac{\sqrt{12 f^2 +9}}{9}\right)}{8 \pi  \sqrt{4  f^2+3}} +O(\rho^4).
  \end{eqnarray}

 \begin{figure}
   \begin{center}
  \includegraphics[width=9cm]{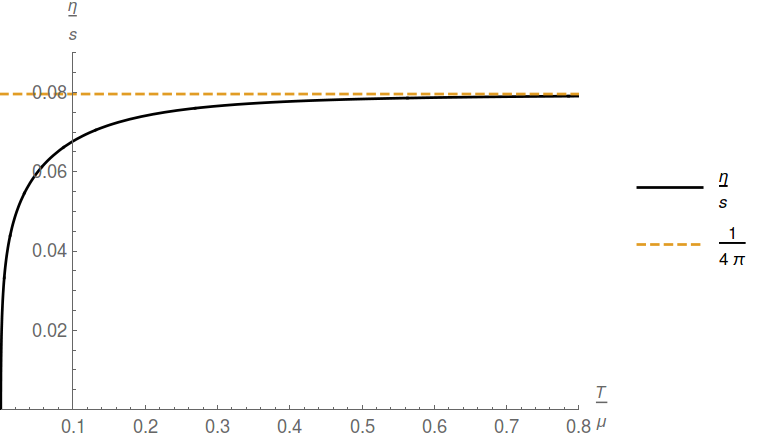}
   \end{center}
    \caption{ Plot of $\eta \over s$ vs temperature T in 5D for $\mu =1$ and $\rho= 0.5$ from Eq.\eqref{5dvis}. } 
    \label{5dvt}
\end{figure}

 \begin{figure}
   \begin{center}
  \includegraphics[width=9cm]{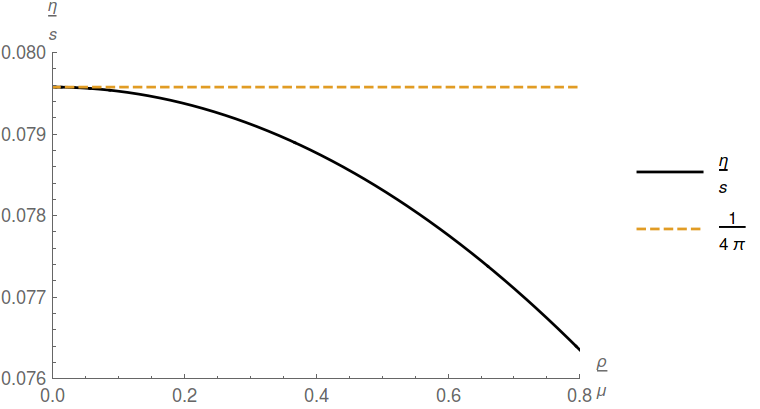}
   \end{center}
    \caption{ Plot of $\eta \over s$ vs anisotropy parameter $\rho$ in 5D for $\mu =1$ and $T= 0.5$ from Eq.\eqref{5dvis}. } 
    \label{5dvr}
\end{figure}

\begin{figure}
   \begin{center}
  \includegraphics[width=9cm]{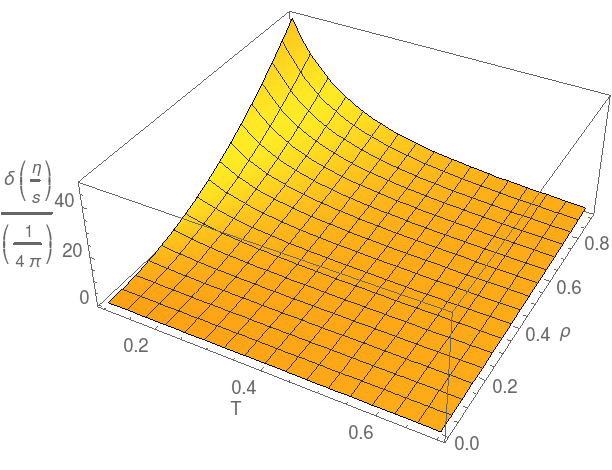}
   \end{center}
    \caption{Percentage reduction of spin 1  shear viscosity over entropy density $ ( \frac{ \delta({ \eta \over s})} { \frac{1}{4 \pi}}$ for the RN anisotropic blackbrane in 5 dimensions for  varying temperature $T \over \mu$ and anisotropy $\rho \over \mu$ (setting $\mu=1$).The reduction is more in the regions of low $T \over \mu$ and high $\rho \over \mu$. } 
    \label{5dred}
\end{figure}
We plot the function ${\eta \over s}$ for varying $\frac{T}{\mu}$ at a fixed anisotropy $\frac{\rho}{\mu} =0.5 $ in Fig.\ref{5dvt}. We find that the violations of KSS bound increase as we decrease ${T \over \mu}$.   Fig.\ref{5dvr} shows  ${\eta \over s}$ for varying $\frac{\rho}{\mu}$ at a fixed temperature $\frac{T}{\mu} =0.5 $. We see that at $\rho=0$ the KSS bound is saturated and is violated at non zero $\rho$. Fig.\ref{5dred} combines the above information in a single 3D plot where we find significant violations of the KSS bound at low ${T \over \mu}$  and high ${\rho \over \mu}$.
\section{Results in 4D dilaton gravity system with finite chemical potential}
\label{setup4d}
 Here we carry out the similar steps for the 4D system consisting of Einstein gravity action in the presence of a massless dilaton and an $U(1)$ gauge field. The blackbrane perturbations we study here has been carried out in details in \cite{Iizuka:2012wt}. Here we wish to be brief and show that the spin 1 components also violate the KSS bound similar to what we found in 5D.  The action is given by;
 \begin{eqnarray}
 S_{bulk}=\int d^4 x\sqrt{-g}\left(R+6\Lambda-\frac{1}{2}\left(\partial\phi\right)^2-\frac{1}{4}F_{\mu\nu}F^{\mu\nu}\right).
 \end{eqnarray}

The 4D metric ansatz that preserves homogeneity but breaks isotropy is given by \footnote{ We have denoted the spatial directions as x and z.};
\begin{eqnarray}
ds^2&=&\frac{L^2}{u^2}\left[-g(u)dv^2-2dvdu+e^{A(u)-B(u)}dx^2+e^{A(u)+B(u)}dz^2\right].
\end{eqnarray}
As discussed in the 5D case, $B(u)$ generates the anisotropy of the metric ansatz with $g(u)>1$ outside the horizon. Setting $g(u)=1$ and $A(u)=B(u)=0$  we recover the $AdS_4$ solution. This corresponds to the case of vanishing dilaton field $\phi$ and gauge field $\mathcal{A}$.

Similar to the 5D case we consider the dilaton to linear in $z$ and the gauge field to be timelike and is a  function of $u$ i.e.
\begin{eqnarray}
\phi &=& \rho z,\\
\mathcal{A}&=&\mathcal{A}_v(u)dv.
\end{eqnarray}

\subsection{4D Reissner Nordstr{\"o}m blackbrane solution}
\label{rn4d}
 To recover the 4D RN blackbrane solution we need to turn off the dilaton  $\phi$. In other words we recover the isotropy in direction $x $ and $z$. This automatically sets $A(u)$ and $B(u)$ to zero. Solving the Einstein's equation and the gauge field equation one can solve for $g(u)$ and $\mathcal{A}_v$ as:
 \begin{eqnarray}
 \mathcal{A}_v&=&(u_h-u){\mu \over u_h},\\
 g(u)&=&c_1 u^3+u^3 \left(\frac{1}{u^3}+\frac{({\mu \over u_h}) ^2
   u}{4}\right).
 \end{eqnarray}
We use the boundary condition that $g(u=u_h)=0$ to solve for $c_1$ as;
\begin{eqnarray}
c_1=-\frac{\mu ^2 u_h^2+4}{4 u_h^3}.
\end{eqnarray}
This gives 
\begin{eqnarray}
g(u)&=&\frac{(u-u_h) \left(\mu ^2 u^3u_h-4
   u^2-4 u u_h-4 u_h^2\right)}{4
   u_h^3}.
\end{eqnarray}
The temperature associated to the outer horizon is given by;
\begin{eqnarray}
T=\frac{12-\mu ^2 u_h^2}{16 \pi  u_h}.
\label{to4d}
\end{eqnarray}
One can solve the above equation to express the outer horizon radius $u_h$ in terms of $T$ and $\mu$ as
\begin{eqnarray}
u_h=\frac{2\left(-4 \pi {T \over \mu} +\sqrt{ 3 + 16 \pi^2 {T^2 \over \mu^2}  }\right)}{\mu}
\end{eqnarray}
Without loss of generality we can consider the inner horizon is located at $u_h/\xi$ where $\xi$ parameter ranging in between $0$ and $1$. Setting $g(u)$ to zero at the inner horizon one can solve for $\mu$ in terms of $u_h$ and $\xi$ as
\begin{eqnarray}
{\mu }=\pm\frac{2 \sqrt{\xi  \left(\xi ^2+\xi
   +1\right)}}{u_h}.
\end{eqnarray}

Plugging this in Eq.\eqref{to4d} we get $T=\frac{3-\xi -\xi^2 -\xi^3}{4 \pi u_h}$.
Demanding $ T>0$ gives ( since $u_h>0$)  $\xi<1$ which is consistent with the definition of inner horizon.
\subsection{Dilaton perturbation to 4D Reissner Nordstr{\"o}m blackbrane}
\label{pert4d}
Now we switch on the dilaton field $\phi(z)=\rho z$ and consider perturbation around the RN blackbrane solution. Let us expand $A(u),B(u)$ and $g(u)$ in power series of $\rho$:
\begin{eqnarray}
A(u)&=&0+\rho^2a_2(u)+\mathcal{O}(\rho^4),\\
B(u)&=&0+\rho^2b_2(u)+\mathcal{O}(\rho^4),\\
g(u)&=&\frac{(u-u_h) \left(\mu ^2 u^3u_h-4
   u^2-4 u u_h-4 u_h^2\right)}{4
   u_h^3}+\rho^2g_2(u)+\mathcal{O}(\rho^4),\\
   \mathcal{A}_v(u)&=&{\mu \over u_h}  \left(u_h-u\right)+\mathcal{O}(\rho^4).
\end{eqnarray}
Substituting these in the Einstein's equation and the gauge field equation, we get
\begin{eqnarray}
a_2(u)&=&c_g,\\
b_2(u)&=&\frac{1}{4 \left(3 \xi ^2+2 \xi +1\right) \sqrt{3 \xi ^2+2
   \xi +3}}\Bigg[-2 (3 \xi +1) u_h^2 \tan ^{-1}\left(\frac{2
   \left(\xi ^2+\xi +1\right) u+(\xi +1)
   u_h}{\sqrt{3 \xi ^2+2 \xi +3}
   u_h}\right)\nonumber\\
   &+&\sqrt{3 \xi ^2+2 \xi +3} \Big(4 c_2 \left(3
   \xi ^2+2 \xi +1\right)+u_h^2 \log
   \left(\left(\xi ^2+\xi +1\right) u^2+(\xi +1) uu_h+u_h^2\right)\nonumber \\
   &-&2 u_h^2 \log(u_h-\xi  u)\Big)\Bigg],\\
g_2(u)&=&-\frac{u^2}{4}+u^3c_m,
\end{eqnarray}
where $c_g,c_2$ and $c_m$ are constants of integration be be determined by appropriate boundary condition. Substituting these in the metric we get;
\begin{eqnarray}
g_{xx}&=&\frac{1}{u^2}\exp\Bigg[c_g\rho^2-\frac{\rho^2}{4 \left(3 \xi ^2+2 \xi +1\right) \sqrt{3 \xi ^2+2 \xi +3}}\Bigg\{\sqrt{3 \xi ^2+2 \xi +3}\Big(4 c_2 \left(3 \xi ^2+2 \xi +1\right)\nonumber \\
&-&u_h^2 \log \left[\left(\xi ^2+\xi +1\right) u^2+(\xi +1)
   u u_h+u_h^2\right]-2 u_h^2 \log
  \left [u_h-\xi  u\right]\Big)\nonumber \\
   &-&2 (3 \xi +1) u_h^2 \tan ^{-1}\left(\frac{2 \left(\xi
   ^2+\xi +1\right) u+(\xi +1) u_h}{\sqrt{3 \xi ^2+2 \xi
   +3} u_h}\right)\Bigg\}\Bigg],\\
g_{zz}&=&\frac{1}{u^2}\exp\Bigg[c_g\rho^2+\frac{\rho^2}{4 \left(3 \xi ^2+2 \xi +1\right) \sqrt{3 \xi ^2+2 \xi +3}}\Bigg\{\sqrt{3 \xi ^2+2 \xi +3}\Big(4 c_2 \left(3 \xi ^2+2 \xi +1\right)\nonumber \\
&-&u_h^2 \log \left[\left(\xi ^2+\xi +1\right) u^2+(\xi +1)
   u u_h+u_h^2\right]-2 u_h^2 \log
  \left [u_h-\xi  u\right]\Big)\nonumber \\
   &-&2 (3 \xi +1) u_h^2 \tan ^{-1}\left(\frac{2 \left(\xi
   ^2+\xi +1\right) u+(\xi +1) u_h}{\sqrt{3 \xi ^2+2 \xi
   +3} u_h}\right)\Bigg\}\Bigg].
\end{eqnarray}
Asymptotically (i.e. $u\to 0$) we expect to get $AdS_4$. Thus expanding $g_{xx}$ and $g_{zz}$ about the asymptotic boundary $u=0$ and setting $g_{xx}=g_{zz}\sim \frac{1}{u^2}$ one can calculate the constants $c_2$ and $c_g$ as
\begin{eqnarray}
c_2&=&-\Bigg[\frac{1}{4 \left(3 \xi ^2+2 \xi +1\right) \left(3 \xi ^2+2 \xi
   +3\right)}\Bigg\{-2 \sqrt{3 \xi ^2+2 \xi +3} u_h^2 \tan
   ^{-1}\left(\frac{\xi +1}{\sqrt{3 \xi ^2+2 \xi
   +3}}\right)\nonumber \\
   &-&6 \xi  \sqrt{3 \xi ^2+2 \xi +3} u_h^2 \tan
   ^{-1}\left(\frac{\xi +1}{\sqrt{3 \xi ^2+2 \xi
   +3}}\right)-6 u_h^2 \log (u_h)-4 \xi  u_h^2 \log (u_h)\nonumber\\
   &-&6 \xi ^2 u_h^2 \log (u_h)+3 \xi ^2
   u_h^2 \log \left(u_h^2\right)+2 \xi 
   u_h^2 \log \left(u_h^2\right)+3
   u_h^2 \log \left(u_h^2\right)\Bigg\}\Bigg],\\
   c_g&=&0.
\end{eqnarray}
\subsection{Spin 1 viscosity in 4D}
\label{vis4d}
\begin{figure}
   \begin{center}
  \includegraphics[width=9cm]{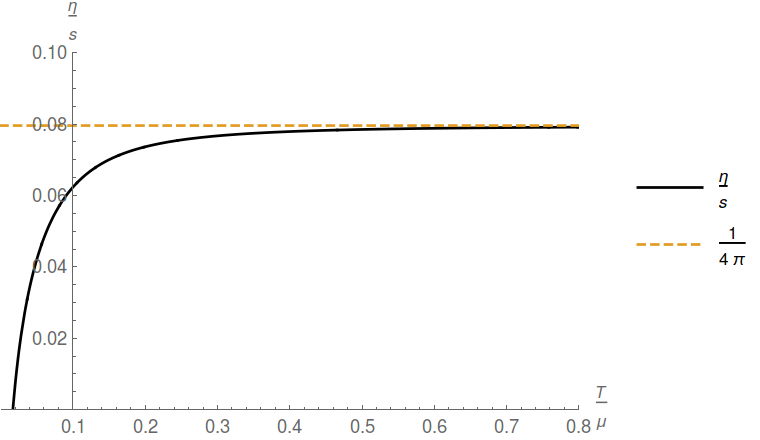}
   \end{center}
    \caption{Plot of $\eta \over s$ vs temperature T in 4D for $\mu =1$ and $\rho= 0.5$ from Eq.\eqref{vis4deq}.}
    \label{4dvt} 
\end{figure}

\begin{figure}
   \begin{center}
  \includegraphics[width=9cm]{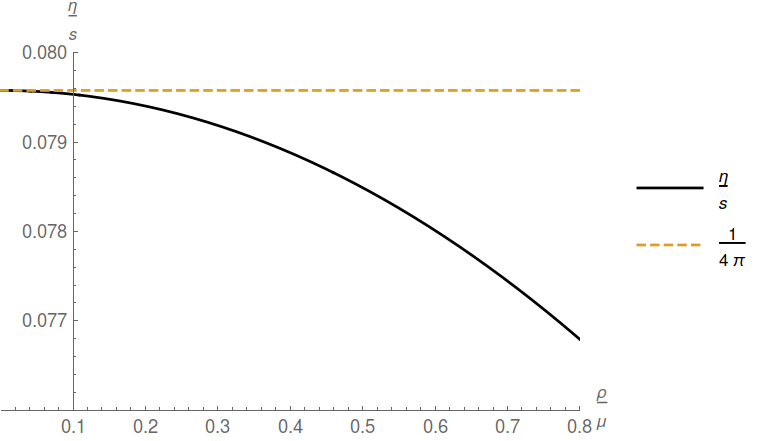}
   \end{center}
    \caption{Plot of $\eta \over s$ vs anisotropy parameter $\rho$ in 4D for $\mu =1$ and $T= 0.5$ from Eq.\eqref{vis4deq}.}
    \label{4dvr} 
\end{figure}

\begin{figure}
   \begin{center}
  \includegraphics[width=9cm]{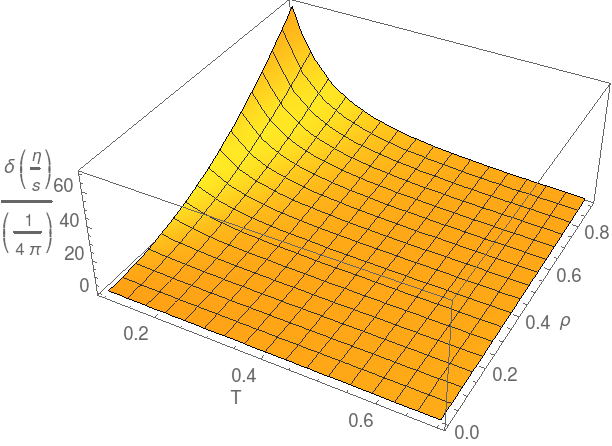}
   \end{center}
    \caption{Percentage reduction of spin 1  shear viscosity over entropy density $ ( \frac{ \delta({ \eta \over s})} { \frac{1}{4 \pi}}$ for the RN anisotropic blackbrane in 4 dimensions for  varying temperature $T \over \mu$ and anisotropy $\rho \over \mu$ (setting $\mu=1$). The reduction is more in the regions of low $T \over \mu$ and high $\rho \over \mu$. } 
    \label{4dred}
\end{figure}

Plugging these constants and expanding $\frac{1}{4\pi}\frac{g_{xx}}{g_{zz}}\big{|}_{u=u_h}$ in $\rho$ about $\rho=0$ gives the viscosity (in units of $s$) as
\begin{eqnarray} \label{vis4deq}
{\eta \over s} &=&\frac{1}{4\pi}\frac{g_{xx}}{g_{zz}}\bigg{|}_{u=u_h}\nonumber \\
&=&\frac{1}{4\pi}-\frac{\rho^2}{8 \pi  \left(3 \xi ^2+2 \xi +1\right) \sqrt{3 \xi ^2+2
   \xi +3}}\Bigg[2 (3 \xi +1) u_h^2 \tan ^{-1}\left(\frac{\xi
   +1}{\sqrt{3 \xi ^2+2 \xi +3}}\right)\nonumber \\
   &-&\sqrt{3 \xi ^2+2 \xi +3} \Big(-u_h^2 \log
   \left(\xi ^2+2 \xi +3\right)+2 u_h^2 \log
   (u_h-\xi  u_h)-2 u_h^2 \log
   u_h\Big)\nonumber \\
   &-&2 (3 \xi +1) u_h^2 \tan ^{-1}\left(\frac{2 \xi
   ^2+3 \xi +3}{\sqrt{3 \xi ^2+2 \xi +3}}\right)\Bigg]+O\left(\rho ^4\right) 
\end{eqnarray}
in the regime  $ \rho \ll (\mu, T)$. In this formula, $\xi$ and $u_h$ are related to T and $\mu$ by the relations ${\mu}=\pm\frac{2 \sqrt{\xi  \left(\xi ^2+\xi
   +1\right)}}{u_h}$ and $T=\frac{12-\mu ^2 u_h^2}{16 \pi  u_h}$ from which we can solve for $u_h$ and $\xi$ in terms of T and $\mu$. 
   \begin{eqnarray}
   u_h&=&\frac{2}{\mu^2}\left(-4 \pi  T+\sqrt{3 \mu ^2+16 \pi ^2 T^2}\right),\\
   \xi&=&-\frac{1}{3}-\frac{2\sqrt[3]{2}}{3}\Bigg[88+\frac{864 \pi ^2 T^2}{\mu ^2}-\frac{216 \pi  T \sqrt{3 \mu ^2+16 \pi ^2T^2}}{\mu ^2}\\ \nonumber
   &+&\sqrt{\left(\frac{864 \pi ^2 T^2}{\mu ^2}-\frac{216 \pi  T \sqrt{3 \mu ^2+16
   \pi ^2 T^2}}{\mu ^2}+88\right)^2+32}\Bigg]^{-\frac{1}{3}}\\ \nonumber
   &+&\frac{1}{3\sqrt[3]{2}}\Bigg[88+\frac{864 \pi ^2 T^2}{\mu ^2}-\frac{216 \pi  T \sqrt{3 \mu ^2+16 \pi ^2T^2}}{\mu ^2}\\ \nonumber
    &+&\sqrt{\left(\frac{864 \pi ^2 T^2}{\mu ^2}-\frac{216 \pi  T \sqrt{3 \mu ^2+16
   \pi ^2 T^2}}{\mu ^2}+88\right)^2+32}\Bigg]^{\frac{1}{3}}.
   \end{eqnarray}
Plugging these in Eq.\eqref{vis4deq} one can get an expression for $\eta/s$ in terms of the parameters T and $\mu$. We plot  ${\eta \over s}$ with respect to $\frac{T}{\mu}$ at a fixed anisotropy $\frac{\rho}{\mu} =0.5 $ in Fig.\ref{4dvt}. Similarly Fig.\ref{4dvr} shows  ${\eta \over s}$ for varying $\frac{\rho}{\mu}$ at a fixed temperature $\frac{T}{\mu} =0.5 $. Fig.\ref{4dred} combines the above information in a single 3D plot where we find significant violations of the KSS bound at low ${T \over \mu}$ and high ${\rho \over \mu}$ just like our results in 5D.

\section{Comparison with the Boltzmann results}
\label{comp}
In general, it is difficult to calculate these corrections due to anisotropy in such strongly coupled systems from first principles since the mean free path is comparable to the
inter-particle separation. The Boltzmann equation is thus inadequate in such regimes. However in the absence of analytical techniques, one can still
estimate these corrections by solving the Boltzmann equation for a system of weakly interacting charged particles in the presence a
constant field in the relaxation time approximation. The relaxation
time $\tau=\lambda/c_s$ where $c_s$ is the speed of sound. $\lambda$ is a parameter which depends on the Fermi energy $E_F=(3\pi^2 n)^{2/3}/(2m)$
and the temperature $T$. The result for a weakly interacting Fermi gas for arbitrary $T/E_F$, is presented in Appendix B of \cite{Samanta:2016pic}. At low $T/E_F$, the
thermal integrals simplify to a great extent and we get
\begin{equation}
\begin{split}
\eta_{xz}=\eta_{yz} &= \eta[1 + 
    c_2 (\lambda k_F)^2({\nabla \phi}/{(\mu k_F)})^2+\calO((\lambda\E/\mu)^4)]
\label{eq:anisotropy_corrections0}
\end{split}
\end{equation}
where $k_F = (3\pi^2 n)^{1/3}$, 
\begin{equation}
\begin{split}
c_2 = -{11}/{28}\;.
~\label{eq:etaoflambda2}
\end{split}
\end{equation}

While we will focus on the spin $1$ component, we note that the corrections to
different components of $\eta$ are different and the shear viscosity tensor is
indeed anisotropic~\cite{Samanta:2016pic}. 

In the absence of potential one obtains the well known result~\cite{Ofengeim:2015qxz}
\begin{equation}
\eta = {k_F^4\lambda}/{(15\pi^2)}~\label{eq:eta0}\;.
\end{equation}
We see that the corrections due to anisotropy are
governed by ${\nabla \phi}\over{(\mu k_F)}$. In the strongly coupled regime $c_2$
(Eq.~\ref{eq:anisotropy_corrections0}) cannot be computed reliably. But it is
intriguing that the weak coupling Boltzmann analysis
(Eq.~\ref{eq:etaoflambda2}) gives $c_2<0$ like the strongly coupled theories
with gravity duals.\\
For example, our calculation in AdS/CFT
theories in the presence of a background chemical potential in the low
anisotropy regime ( $\rho \ll (\mu, T)$ ) on the gravity side shows that 
the coefficient $c_2$ is negative even in the presence of a finite chemical
potential. The formula Eq.\ref{5dvis}  gives $c_2=-0.094$ at ${T \over \mu}=0.4$.\\

\section{Conclusions}
Let us conclude by summarizing the results and some open questions we will like to address in future. We considered a system which is the standard Einstein Maxwell theory with a negative cosmological constant. This system is known to admit the Reissner Nordstr{\"o}m Blackbrane solution. We next introduce a small dilaton that breaks rotational symmetry. We find the blackbrane solution perturbatively in the dilaton anisotropy parameter in 5 dimensions. This blackbrane solution corresponds to an anisotropic phase at finite tempearture and chemical potential where the anisotropy is small compared to the other two mass scales, $\mu$ and T.  We find that in this simple setup the  components of the anisotropic shear viscosity tensor, which are spin one with respect to the surviving symmetry after breaking of rotational invariance, violates the KSS bound ${\eta \over s}\ge {1 \over 4 \pi} $. Similar violations were also reported in \cite{Ge:2014aza} where some components of shear viscosity violated the bound in the Einstein-Maxwell-dilaton-axion theory. In our computation, we used the results of \cite{Jain:2015txa} where it was shown that in all situations where the rotational symmetry is broken by a constant driving force, there exists a general formula for the shear viscosity to entropy density for these spin one components in terms of the ratio of the metric components evaluated at the horizon.\\
We compare our results with that of a system of weakly interacting charged fermions in an external field in Sec.\ref{comp}. It is quite surprising that both the weakly coupled system as well as the strongly coupled theories with smooth gravity duals predict a reduction in $\eta/s$ from the KSS value. Our results also open up an exciting possibility to experimentally measure such violations in ultracold fermi gases at unitarity as explained in detail in \cite{Samanta:2016pic,Samanta:2016lsh}. Fig.\ref{5dred} and Fig.\ref{4dred} suggest that significant violations of the KSS bound may be found at low ${T \over \mu}$ and strong anisotropy ${\rho \over \mu}$ in such systems.\\
In this paper, we explored a situation where we have small anisotropy at a finite chemical potential. One can also look at the other limit ie. adding a small chemical potential to an anisotropic background.  It will be interesting to study transport in this regime.
Another issue to investigate is the stability of these solutions. This requires a quasinormal mode analysis similar to that performed in \cite{Jain:2014vka}. We hope to report on these in future. \\

\acknowledgments

The authors would like to thank the organizers of the
34th Jerusalem Winter
School on Theoretical Physics ``New Horizons in Quantum Matter"  at The Hebrew University, where some of this work was carried out. We also thank Sandip Trivedi, Rishi Sharma, Sachin Jain and Nilay Kundu for discussions. 

\appendix
\section{Conductivity from shear viscosity }
\label{condres}
In this Appendix, we show how the spin 1 viscosity components in (d+1) dimensions can be related to conductivity in d dimensions. Using this relation, we derive the ratio of conductivity over susceptibility found by Kovtun and Ritz in \cite{Kovtun:2008kx}. The basic idea behind the analysis is dimensional reduction which maps the gravitational spin 1 shear modes in the higher dimensional theory to a gauge field in the lower dimensional theory once we compactify along one of the spatial directions. Here we restrict ourselves to the simple case where we have an isotropic background solution in the higher dimensional theory.
\subsection{The Dimensionally Reduced Theory}

To be concrete,  let us start in  $5$ dimensions with a gravitational action :

\be \label{kkh}
S=\frac{1}{2\hat{\kappa}^2}\int d^{5}x\sqrt{-\hat {g}}~(\hat{R} + 12 \Lambda) .
\ee
Here $2\hat{\kappa}^2=16\pi \hat{G}$ where $\hat{G}$ is the Newton's Constant in 5-dimension and let $\Lambda$=1.\\
Let us parametrise the $5$ dimensional metric as follows 
\begin{equation}
\left( \hat{g}_{AB} \right) = \left( \begin{array}{cc}
    e^{-\psi(u)} g_{\mu\nu} +  e^{2 \psi(u)}  A_{\mu} A_{\nu} \; \; & \; \; 
       e^{2 \psi(u)}  A_{\mu} \\
    e^{2 \psi(u)}  A_{\nu} \; \;                                & \; \; 
      e^{2 \psi(u)}
   \end{array} \right) \; \; \; ,
\label{5dMetric}
\end{equation}
For simplicity, we take assume all metric  components to be independent of the $z$ direction along which we wish to compactify. Hatted metric components belong to the higher dimensional theory. After starightforward KK reduction, we get\\
\begin{eqnarray}
\label{dimredac}
S =\frac{ 1}{2\kappa^2}\int d^{4}x \sqrt{-g} \left( R - {3 \over 2}(\partial\psi)^2  - {e^{3 \psi}\over 4} F^{2} +12 e^{- \psi} \right),
\end{eqnarray}
where we have dropped total derivatives .  \\
In this parametrisation,
\be 
\label{gzz}
\hat{g}_{zz}=e^{2\psi}.
\ee

${\kappa}$  and ${\hat{\kappa}}$ are related by 
\be
\label{gn0}
{L \over 2 \hat{\kappa}^2} = {1 \over 2 \kappa^2 },
\ee
where 
$L$   is the length of the compact $z$ direction. 

We thus get an effective low dimensional action for the gauge field
\be
\label{genac}
S_{Gauge} =\frac{1}{2\kappa^2} \int d^4 x \sqrt{-g} \left( {- 1 \over 4 g_{\rm{eff}}^{2}(u)} F^2  + other ~ terms \right),
\ee
where
\be
\label{effc}
{ 1 \over  g_{\rm{eff}}^{2}(u)} = e^{3 \psi}=\left(\hat{g}_{zz}(u)\right)^{3\over2}.
\ee

Such a system was also considered in \cite{Myers:2009ij,Chakrabarti:2010xy}. 
  To study the conductivity we consider a  perturbation for the $x$ component of the gauge field,
\begin{equation}
\label{gans}
{ A}_{x} (\vec x,t,u) = \int{ d\omega d^{3}\vec k \over (2 \pi)^{4} } e^{-i \omega t + \vec k.\vec x}Z(u,\omega) .\quad
\end{equation}
This gauge field  perturbation decouples from the rest and
 $Z(u,\omega)$ satisfies the equation
\begin{equation}
\frac{d}{du}(N(u) \frac{d}{du}Z(u,\omega))-\omega^2 N(u)~ g_{uu} g^{tt} Z(u,\omega)=0,
\label{eqnmotion}
\end{equation}
 
with 
\be
\label{defN}
N(u) = \sqrt{-g}\frac{1}{g_{\rm{eff}}^2}g^{xx}g^{uu},
\ee

Following the steps in \cite{Jain:2015txa}
\be
\label{fsigma}
\sigma = \frac {1}{2 \kappa^{2}} {\hat{g}_{xx}^{2} \over \sqrt{\hat{g}_{zz}}}.
\ee

   The next step is to relate the conductivity obtained above to the viscosity. For the spin 1 viscosity components, the connection is straightforward and discussed in detail in Sec.5.2 of \cite{Jain:2015txa}. 

   \be
   \label{viscon}
   \eta_{xz}= {\sigma \over L}.
   \ee
   The entropy density in the $5$ dimensional theory is given by 
 \be
 \label{entd0} 
 s={2\pi\over \hat{\kappa}^{2}} A= {2\pi\over \hat{\kappa}^{2}} \sqrt{\hat{g}_{xx} \hat{g}_{yy}\hat{g}_{zz}},
 \ee

Let us now briefly review the results of Kovtun and Ritz  \cite{Kovtun:2008kx} . We refer to  \cite{Kovtun:2008kx} for additional details.
We consider field theories with classical gravity duals in Anti-de Sitter space.
We consider the RN black hole solution in AdS in a theory given by the following action 
\begin{equation}
  S = \frac{1}{16\pi G} \int\!\! {\rm d}^{d+1}x\, \sqrt{-g}\;
      \left[R+\frac{d(d{-}1)}{L^2}\right] 
     -\frac{1}{4g_{d+1}^2} \int\!\! {\rm d}^{d+1}x\, \sqrt{-g}\;
      F^2\,,
\label{einmax}
\end{equation}
where $L^2$ denotes the cosmological constant,
and $g_{d+1}^2$ is the gauge coupling .

The susceptibility  is found as

\begin{equation}
 \chi = \frac{(d{-}2)L^{d-3}}{g_{d+1}^2}
        \left(\frac{4\pi}{d}\right)^{d-2} T^{d-2}.
\label{eq:chi}
\end{equation} 
where T denotes the temperature of the RN black hole. The above formula can also be written as
$ \chi= (e^2/g_{d+1}^2)\, (L/z_0)^{d-3} (d{-}2)/z_0 $ changing the radial coordinate to $z=L^2/r$ and the CFT temperature is $T= d/(4\pi z_0)$ where $z_0$ is the outer horizon radius of the RN balckhole.

Now ,
\be
\frac{\sigma}{\chi }=\frac{\sigma }{S_{higher}}\frac{S_{higher}}{\chi }=\frac{\eta L  }{S_{higher}}\frac{S_{higher}}{\chi }=\frac{L}{4 \pi}\frac{\frac{1}{4G}\frac{\sqrt{g}}{\sqrt{g_{tt}g_{uu}}}|_{z=z_0}}{\frac{1}{g_{d+1}^{2}}( \frac{L}{z_{0}})^{d-3}(\frac{d-2}{z_{0}})}
\ee
where $S_{higher}$ denotes the higher dimensional entropy density. Here we used the results of \cite{Jain:2015txa} ie. eq.\ref{viscon} to relate the conductivity to shear viscosity and used the fact the $\frac{\eta}{s} = \frac{1}{4\pi}$ for the higher dimensional isotropic background geometry.

Comparing the action eq.(\ref{einmax})  with the dimensionally reduced action in eq.(\ref{genac})
\be
L e ^{3 \psi}=\frac{16 \pi G}{g_{d+1}^{2}}
\ee
Using this above , we find 
\be
\frac{\sigma}{\chi }=\frac{L}{4 \pi}\frac{\frac{1}{4G}\frac{\sqrt{g}}{\sqrt{g_{tt}g_{uu}}}|_{z=z_0}}{\frac{L e ^{3 \psi}}{16 \pi G}( \frac{L}{z_{0}})^{d-3}(\frac{d-2}{z_{0}})}
\ee
Assuming isotropicity in higher dimension $ e ^{ \psi}|_{z=z_{0}}=\frac{L}{z_{0}}$ we arrive at the result
\be
\frac{\sigma}{\chi }=\frac{(\frac{L}{z_{0}})^{d}}{( \frac{L}{z_{0}})^{d}\frac{d-2}{z_{0}}}=\frac{z_{0}}{d-2}=\frac{1}{4 \pi T}\frac{d}{d-2}
\ee
found by Kovtun and Ritz in \cite{Kovtun:2008kx}. Thus the spin one viscosity components are related to conductivity in a lower dimensional theory \footnote{This observation was made in collaboration with Sachin Jain}.



\bibliographystyle{ieeetr}
\bibliography{ucver1}

\end{document}